\documentclass[aip,graphicx,reprint]{revtex4-1}
\usepackage{graphicx}
\usepackage{amsmath}
\usepackage{amssymb}

\begin{document}


\title{Large-Scale Parallel Electric Fields and Return Currents in a Global Simulation Model}
\author{H. Arnold}
\affiliation{IREAP, University of Maryland, College Park MD 20742-3511, USA}
\author{J.~F.~Drake}
\affiliation{IREAP, University of Maryland, College Park MD 20742-3511, USA}
\author{M.~Swisdak}

\affiliation{IREAP, University of Maryland, College Park MD 20742-3511, USA}
\author{J. Dahlin}
\affiliation{NASA Goddard Space Flight Center}

\date{\today}

\begin{abstract}
A new computational model, {\it kglobal}, is being developed to
explore energetic electron production via magnetic reconnection in
macroscale systems. The model is based on the discovery that the
production of energetic electrons during reconnection is controlled by
Fermi reflection in large-scale magnetic fields and not by parallel
electric fields localized in kinetic scale boundary layers. Thus, the
model eliminates these boundary layers. However, although the parallel
electric fields that develop around the magnetic x-line and
associated separatrices are not important in producing energetic
electrons, there is a large scale electric field that kickstarts the
heating of low-energy electrons and drives the cold-electron return
current that accompanies escaping energetic electrons in open
systems. This macroscale electric field is produced by
magnetic-field-aligned gradients in the electron pressure. We have
upgraded {\it kglobal} to include this large-scale electric field
while maintaining energy conservation. The new model is tested by
exploring the dynamics of electron acoustic modes which develop as a
consequence of the presence of two electron species: hot kinetic and
cold fluid electrons. Remarkably, the damping of electron acoustic
modes is accurately captured by {\it kglobal}. Additionally, it has been established that {\it kglobal}
correctly describes the dynamics of the interaction of the parallel
electric field with escaping hot electrons through
benchmarking simulations with the Particle-In-Cell (PIC) code
{\it p3d}.
\end{abstract}


\maketitle 

\section{Introduction}
Solar flares convert magnetic energy into particle energy in the solar
corona via magnetic reconnection
\citep{Lin1971,Emslie2004,Emslie2005}. From observations, we know that
a significant fraction of the released energy during the reconnection
event can go into accelerating energetic electrons, which form a
suprathermal tail that takes the form of a power law
distribution\citep{Heristchi1992,Emslie2004,Emslie2012}. The plasma
pressure from these nonthermal particles can be comparable to the
pressure of the ambient magnetic field
\citep{Krucker2010,Oka2013}. Further, using observations from the Wind spacecraft in the
distant magnetotail, \citet{Oieroset2002} found energetic electrons in a
broad region around the x-line rather than in narrow boundary layers
that would be expected in laminar 2D reconnection models \citep{Dahlin2017,Drake2005}. Additionally,
observations from the Reuven Ramaty High Energy Solar
Spectroscopic Imager (RHESSI) and the Atmospheric Imaging Assembly on
the Solar Dynamic Observatory, \citet{Krucker2014} revealed that a large
fraction of the electron population in the above-the-loop-top source
was energized to suprathermal energies. Multiple x-line reconnection
can potentially describe both the diffuse distribution of energetic
electrons seen in the \citet{Oieroset2002} and \citet{Krucker2014} papers, and
the large number of energetic electrons seen in flares \citep{Miller1997,Emslie2012}. This is
because multiple x-line reconnection in 3D is turbulent and enables
electrons to undergo acceleration in a much larger volume than in a 2D system \citep{Dahlin2017}. Additionally, it is
now well known that current sheets can spawn multiple, volume-filling x-lines in three-dimensional systems in the presence of a guide field \citep{Drake2006,Daughton2011,Dahlin2015,Dahlin2017}. This behavior does not take place in anti-parallel reconnection, even in a 3D system \citep{Dahlin2017}.

In the past, simulations that study flares have been based on PIC
codes, magnetohydrodynamic (MHD) codes, or some combination thereof
such as hybrid codes or MHD codes with an embedded PIC region
\citep{Gordovskyy2019}. However, explicit PIC models require that
kinetic scales such as the Debye length be resolved. The Debye length
is typically less than a centimeter in the solar corona yet the size
of the flare itself can be up to ten orders of magnitude larger, a typical
size being $10^4$ km. Thus, using a PIC code to fully model a solar
flare is computationally impossible. An MHD model, on the other hand, is not constrained
by the need to resolve kinetic scales. However, since there are no
particles in an MHD model, studying particle acceleration is not
possible except by exploring the motion of test particles. While test
particles can illuminate certain aspects of particle acceleration in
solar flares, such as the primary mechanisms responsible for
acceleration, there is no feedback of the energetic particles on the
fields \citep{Guidoni2016}. The particle energy can therefore run
away. An embedded PIC code is an alternative but, since the energy
release volume in flares is large and the region the particles are
accelerated is also broadly distributed, the separation of scales
problem still exists. There are two dominant mechanisms that are responsible for
particle heating and acceleration in magnetic reconnection: direct
acceleration from parallel electric fields in diffusion regions and
along magnetic separatrices \citep{Ball2018,Wang2016}, and Fermi acceleration. Of these two, the
latter is large scale and does not require resolving kinetic
scales. Additionally, it has recently been proposed that Fermi
acceleration, which occurs on macro scales, is what drives non thermal
particles and hence contributes to particle energization, rather than
exclusively contributing to heating \citep{Alves2018,Dahlin2016, Drake2006b,
  Drake2013, Guo2019, Li2017, Li2018}. Thus, we are developing a new computational model, {\it
  kglobal}, that takes advantage of this discovery to order out all
kinetic scales that must be resolved in PIC codes and conserves energy
by particle feedback on the fields \citep{Drake2019}. This model has
an MHD backbone, but also includes self-consistent feedback from
particle electrons.

The process of magnetic reconnection leads to the formation of bent
field lines whose tension drives an exhaust travelling at the upstream
Alfv\'en speed \citep{Lin1993}. Particles stream in along these bent,
reconnected field lines which act like a moving wall and thus "kick"
the particles which then increase in speed by twice the Alv\'en
speed \citep{Drake2006b}. However, it is known that a large-scale
parallel electric field (not localized in boundary layers) facilitates
this process by confining electrons within the reconnection exhaust
such that they undergo multiple Fermi kicks \citep{Egedal2008,
  Egedal2012, Egedal2015, Haggerty2015}. Thus, it is of interest to include this
large scale parallel electric field in our model to properly model the
energy gain of low energy electrons. This potential is not,
however, important to the dynamics of very energetic electrons. This
field arises from parallel gradients in the electron pressure and
points away from the current sheet in the reconnection exhaust. In an
open system it then drives a return current of cold electrons that
balances the current associated with escaping hot electrons to
maintain zero net parallel current. We have updated the computational
model {\it kglobal} to include the large scale electric field and
present the results of testing herein. See the Appendix for a calculation of the parallel electric field and the exploration of energy conservation when this field is included.

\section*{The {\it kglobal} model with $E_{||}$}
Since the parallel electric fields that develop in kinetic scale
boundary layers \citep{Drake2005, Pritchett2004} are ineffective
drivers of energetic electrons during reconnection
\citep{Dahlin2014,Dahlin2016,Dahlin2017}, we have formulated a model in which all
kinetic scale boundary layers are eliminated \citep{Drake2019}. This
new model includes the key physics necessary to produce high energy
particles without having to resolve kinetic scales. We do this by
representing hot electrons as particles and cold electrons and ions as an
MHD fluid. The hot electrons are evolved using the guiding center
equations and they feed back on the fluid through their gyrotropic
pressure tensor in the ion momentum equation. The electric and
magnetic fields are evolved in the usual way from the MHD
fluid. \citet{Drake2019} presented in detail the derivation of this
model.  Crucially, this model conserves energy, which prevents the
electron energy from running away. The dominant feedback is through
the development of pressure anisotropy of the energetic electrons -- a
strong increase of the parallel electron pressure weakens the magnetic
tension that drives reconnection, thereby throttling magnetic energy
release. MHD codes are able to achieve normalized rates of reconnection that are of the order of $0.01$ through the formation of multiple plasmoids. This rate is smaller than typical rates from PIC simulations \citep{Liu2017,Shay2007}. However, through the introduction of artificial resistivity and hyperviscosity fast rates of reconnection can be achieved in the MHD model \citep{Uzdensky2010}. Care must be taken, however, that artifical dissipation does not suppress multi, x-line reconnection, which is required to produce a non-thermal particle spectrum. Our plan is to explore various approaches to achieve fast reconnection while minimizing the impact on multi x-line formation. We should be able to correctly capture the physics of the
acceleration of suprathermal electrons in a macroscale system with none of the constraints associated with including
kinetic-scale boundary layers -- there are no kinetic-scale boundary
layers in the model. This {\it kglobal} code is operational and
preliminary tests of its capabilities have been described in
\citet{Drake2019}. It correctly describes an Alfv\'en wave in the
presence of a pressure anisotropy and reproduces the linear growth
rate of the firehose instability.

The large-scale parallel electric field is obtained by combining the
parallel momentum equations for the three species (ions, cold
electrons and hot electrons) into a single equation for the total
parallel current. Because of constraints on this current, the driver
of the current must be small and therefore can be set to zero, which
yields a constraint equation for the parallel electric field. The
details of the calculation are shown in the Appendix. The resulting
expression for the parallel electric field is given by
\begin{equation}\label{epar}
E_{\parallel}=\frac{-1}{n_ie}\left( \boldsymbol{B} \cdot \boldsymbol{\nabla} \left( \frac{m_en_c v_{\parallel c}^2}{B} \right) + \boldsymbol{b}\cdot \boldsymbol{\nabla} P_{c} + \boldsymbol{b} \cdot \boldsymbol{\nabla} \cdot \boldsymbol{T}_h\right)
\end{equation}
where $m_e$ is the electron mass, $n_c$, $n_h$, $n_i=n_c+n_h$, $v_{||c}$, $v_{\parallel h}$, and $v_{\parallel i}$ are the densities and flow speeds (parallel to the magnetic field) of the two electron species and the ions respectively, $P_c$ is the scalar pressure of the cold electron fluid, $B$ is the magnetic field, $\boldsymbol{ b}$ is a unit vector along $\boldsymbol{B}$, and $\boldsymbol{T}_h$
is the gyrotropic stress tensor of the hot electron particles,
including their inertial contributions \citep{Drake2019},
\begin{equation}
  \boldsymbol{T}_{h}=T_{eh\parallel}\boldsymbol{bb}+P_{eh\perp}(\boldsymbol{I}-\boldsymbol{bb}),
  \label{eqn:T_h}
\end{equation}
where $\boldsymbol{I}$ is the unit tensor, $T_{eh\parallel}$ is the stress
tensor along the magnetic field $\boldsymbol{B}$ and $P_{eh\perp}$ is the
usual perpendicular pressure,
\begin{equation}
  P_{eh\perp}=\int d\boldsymbol{p}_{e}\frac{p_{e\perp}^2}{2m_e \gamma_e}f,
    \label{eqn:Pperp}
\end{equation}
where in the frame drifting with $\boldsymbol{v_E}=c\boldsymbol{E}\times \boldsymbol{B}/B^2$ there are no perpendicular flows so
$f=f(\boldsymbol{x},p_{e\parallel}, p_{e\perp},t)$. $T_{eh\parallel}$ includes
the mean parallel drifts of the hot electrons,
 \begin{equation}
  T_{eh\parallel}=\int d\boldsymbol{p}_{e}\frac{p_{e\parallel}^2}{m_e \gamma_e}f,
  \label{eqn:Tpar}
 \end{equation}
 with $p_{e\parallel}$ the hot parallel electron momentum with relativistic
 factor $\gamma_e$. The normalizations for {\it kglobal} described in
 \citet{Drake2019} remain unchanged. However, we now have a separate
 normalization for the parallel electric field, $E_\parallel\sim
 m_eC_{Ae}^2/eL_0=m_iC_A^2/eL_0$ where $C_{Ae}$ is the electron
 Alfv\'en speed, and $L_0$ is the length scale of the domain. The normalization for
 $E_\parallel$ comes from parallel force balance. Compared with the
 usual scaling for the perpendicular electric field $\boldsymbol{E}_\perp\sim
 C_AB_0/c$, the parallel electric field satisfies $E_\parallel/E_\perp \sim d_i/L_0\ll 1$. Thus we only keep the parallel electric field for motion along the field lines and it can therefore be neglected
 in Faraday's equation when evolving the magnetic field. The addition of this electric field modifies the momentum equation for the ions and the guiding center equation for the particle electrons from \citet{Drake2019} in the following way:
\begin{multline}
\rho\frac{d\boldsymbol{v}}{dt}=\frac{1}{c}\boldsymbol{J}\times\boldsymbol{B}-\boldsymbol{\nabla}P_i-\boldsymbol{\nabla}_{\perp} P_c-( \boldsymbol{\nabla}\cdot\boldsymbol{T}_{eh})_{\perp} + en_iE_{||} \boldsymbol{b} \\- m_en_cv_{||c}^2 \boldsymbol{\kappa}
\label{fluidmom}
\end{multline}
\begin{equation}
  \frac{d}{dt}p_{e||}=p_{e||} \boldsymbol{v_E} \cdot \boldsymbol{ \kappa} - \frac{\mu_e}{\gamma_e} \boldsymbol{b} \cdot \boldsymbol{\nabla} B - eE_{||}
  \label{eqn:ehot}
\end{equation}
where $\kappa = \boldsymbol{b} \cdot \boldsymbol{\nabla} \boldsymbol{b}$ is the magnetic curvature and $\mu_e=p_{e \perp}^2/2m_eB$ is the magnetic moment of the electron. Note that in Eq.~(\ref{fluidmom}) the gradients of the cold electron pressure and hot electron stress tensor are now in the perpendicular direction only. See the Appendix for a derivation of Eq.~(\ref{fluidmom}). Since the parallel electric field is the same order as the pressure terms in Eq.~(\ref{fluidmom}), thermal particles are reflected by this electric potential, which prevents heated electrons from escaping from the reconnection diffusion region and the exhaust \citep{Egedal2012, Haggerty2015}. The consequence for electrons is that they can undergo multiple Fermi reflections within the reconnection exhaust, which facilitates the initial energy gain of electrons.

With the inclusion of a large-scale parallel electric field, {\it
  kglobal} should correctly describe the dynamics of hot electrons
escaping along the ambient magnetic field in an open system and the
development of a return current of cold electrons. The large-scale
parallel electric field suppresses the escape of hot electrons and
drives a return current of cold electrons. In its most basic form
this dynamic can be reduced to that of an electron acoustic mode, which
can exist in plasmas with separate and distinct electron
populations \citep{Gary1985}. In the electron acoustic mode the
electrons slosh back and forth on a short time scale so that the ions
are practically stationary. Thus, we benchmark {\it kglobal} by
simulating this process.

\section{Testing}
\begin{figure}
\centering
\includegraphics[width=16pc,height=8pc]{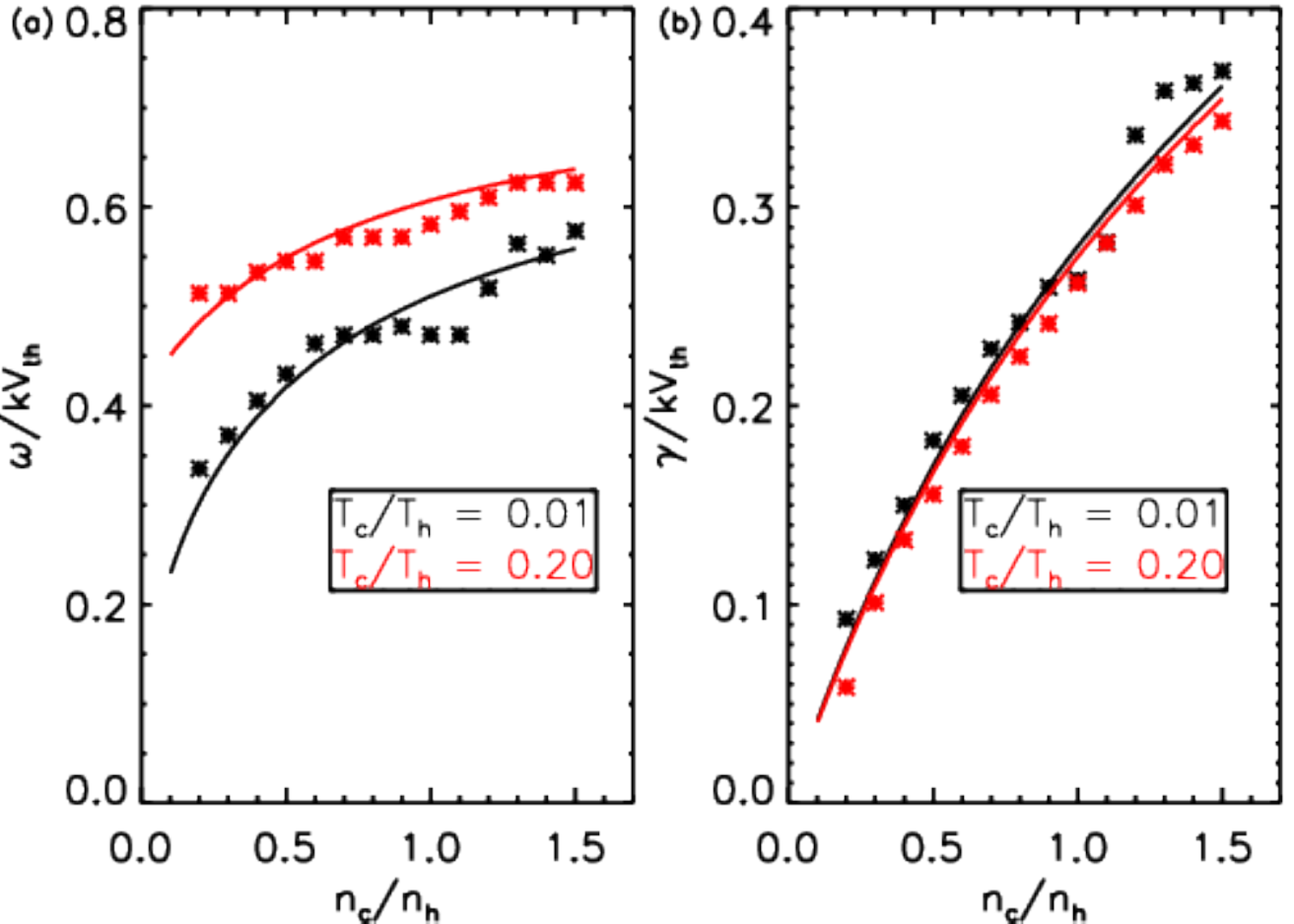}
\caption{In panel (a) the electron acoustic wave phase speed versus the cold to hot density ratio of the electrons. In panel (b) The electron acoustic wave damping rate versus the cold to hot density ratio of the electrons. The stars are taken from the simulations and the lines are from the linear theory. Note that the phase speed is normalized to the thermal speed of the hot electrons and the damping rate is normalized to the time a thermal particle requires to travel one wavelength.}
\label{acoustic}
\end{figure}

Since electron acoustic waves only involve electron motion parallel to the magnetic field, the only non-zero
gradients are along the magnetic field. Thus, the perturbed distribution function,
 $\tilde{f}$, of the hot electrons is only a
function of $v_{||}$ and $x_{||}$. We obtain
\begin{equation}
  \partial_t \tilde{f} + v_{||} \boldsymbol{\nabla_{||}} \tilde{f} - \frac{e}{m_e} \tilde{E_{||}} \partial_{v_{||}} f_0 = 0.
  \label{tildef}
\end{equation}
Similarly, by enforcing charge neutrality and taking the cold electron pressure from the constancy of $P_{c}/n_{c}^{5/3}$ , Eq.~(\ref{epar}) becomes
\begin{equation}
\tilde{E}_\parallel=-\frac{1}{n_ie} \left( \frac{5}{3} \boldsymbol{\nabla}_\parallel T_{c}\tilde{n}_{h} + \boldsymbol{\nabla_\parallel} \tilde{T}_{h} \right). 
\end{equation}
By assuming that the unperturbed hot electron distribution function is a Maxwellian, we can solve Eq.~(\ref{tildef}) for $\tilde{f}$ and take the moments  to obtain the first order corrections to the hot electron density and pressure. After some algebra the dispersion function for the electron acoustic wave is:
\begin{equation}
  \frac{n_{0c}}{n_{0h}}=Z'(\zeta) \left(\frac{5}{6}\frac{T_{0c}}{T_{0h}}+\zeta^2 \right)
  \label{dispersion}
\end{equation}
where $n_{0c}$ is the unperturbed density of the cold electrons
(fluid), $n_{0h}$ is the unperturbed density of the hot electrons
(particles), $T_{0c}$ is the unperturbed temperature of the cold
electrons, $T_{0h}$ is the unperturbed temperature of the hot
electrons, $\zeta=\omega/kv_{th}$, $v_{th}$ is the thermal speed of
the hot electrons, and $Z'(\zeta)$ is the derivative of the plasma
dispersion function. Note that this result matches that of
\citet{Gary1985} in the long wavelength limit $k\ll k_{De}$ where
$k_{De}^{-1}$ is the Debye length. For $T_{0c}\ll T_{0h}$ and
$n_{0c}\ll n_{0h}$ the phase speed of the wave is small compared with
$v_{th}$ and the mode is only weakly damped and has a characteristic
frequency
\begin{equation}
  \omega=kv_{th}\sqrt{\frac{n_{0c}}{n_{0h}}+\frac{5}{6}\frac{T_{0c}}{T_{0H}}}.
\end{equation}
We numerically solved this equation for various values of the density
and temperature ratios and obtained the frequency and decay rates of
these waves. For each value of the two parameters, we initialized {\it
  kglobal} with a sinusoidal perturbation in the electron density and
temperature and measured the corresponding frequencies and decay rates
of the resulting disturbance. The results of the linear theory and the
simulation results are plotted in Fig.~\ref{acoustic}. The damping
rate of the mode is controlled by the Landau resonance with the
energetic component which is accurately captured by the code, a
remarkable result. A similar argument can show that {\it kglobal} can damp ion acoustic waves with Landau damping as well.

\begin{figure}
    \centering
    \includegraphics[width=16pc]{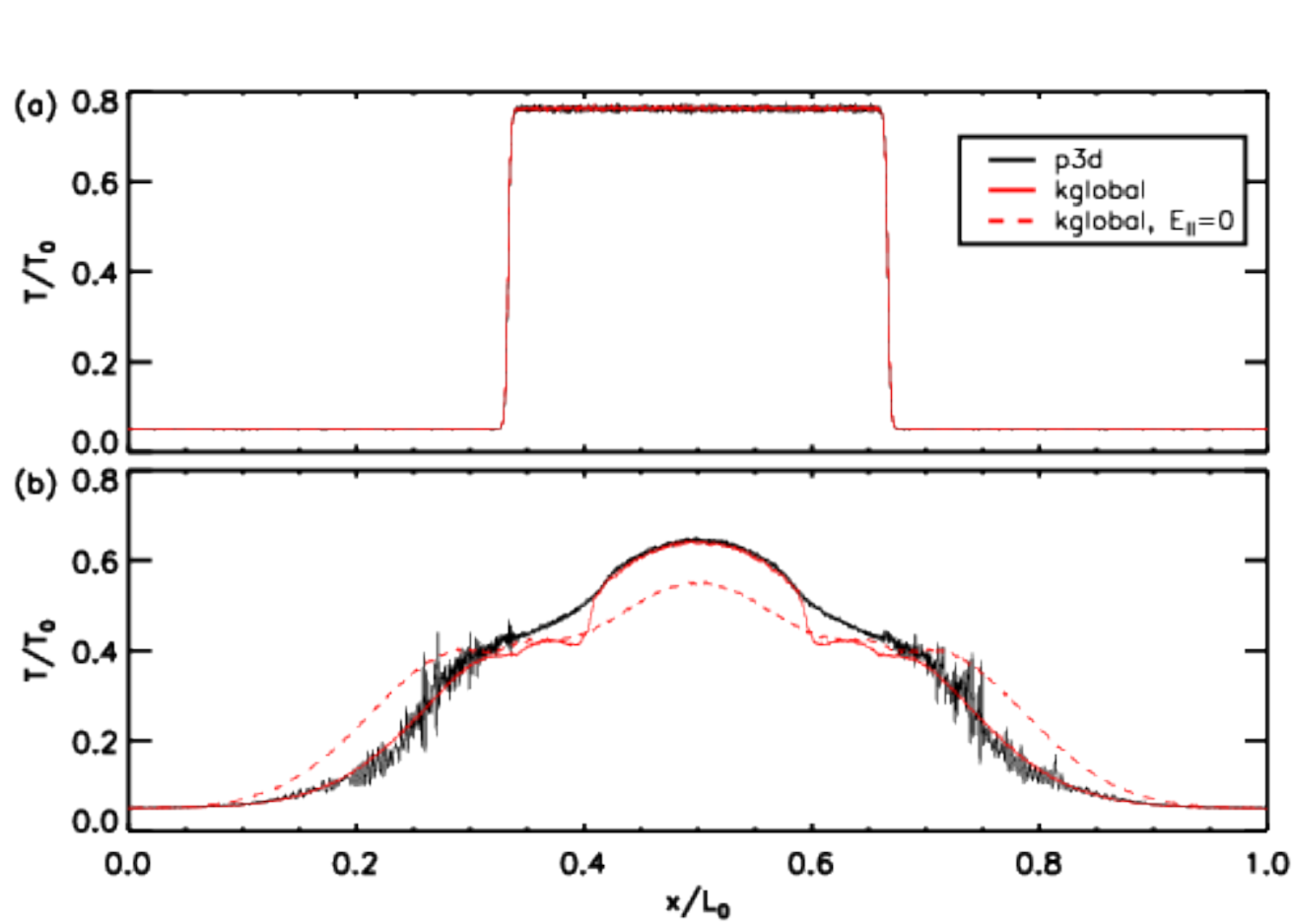}
    \caption{Profiles of the total electron temperature for the PIC code {\it p3d} (black), the {\it kglobal} code with the included parallel electric field (red) and without the parallel electric field (dashed red). In panel (a) at $t/t_{th}=0$. In panel (b) at $t/t_{th}=0.12$ where $t_{th}$ is the time a hot thermal electron requires to travel the length of the box.}
    \label{tpar}
\end{figure}

In our final test we compare a simulation with {\it kglobal} to a
simulation with the PIC code {\it p3d} \citep{Zeiler2002}. We set up a
simplified version of what we expect to see in a reconnection
exhaust. The initial conditions consist of a constant magnetic field,
a constant density made up of 75\% particle electrons and 25\% fluid
electrons, and a temperature profile for the particle electrons that increases sharply in the
center to twenty times the asymptotic value as can be seen in
Fig.~\ref{tpar}(a). This value of the hot to cold electron density
ratio was chosen to quicken the dynamics since we know from
Fig.~\ref{acoustic}(b) that the larger the ratio the larger the
damping rate. To convert this setup to a PIC version, we had to make
sure that the smallest length scale in {\it kglobal} was much larger
than the Debye length since this scale is not resolved in {\it
  kglobal}. Thus we equated the transition width between the two
regions of hot and cold electrons to $30$ times the Debye length.

In both simulations we utilized a large spatial domain in the parallel
direction so there is space for the hot electrons to expand. A small
domain in the perpendicular direction was included so that the data
could be averaged over this direction to decrease particle noise. For
{\it kglobal} we had a domain of 2048 x 64 cells and for {\it p3d}, 8192 x
64. In both simulations the electron to ion mass ratio was $1/1836$
and the speed of light was $300$ times the Alfv\'en speed. For {\it p3d} a
uniform background with constant density and a temperature
corresponding to the cold electron fluid in {\it kglobal} was included
along with an electron population with the same temperature profile as the hot species in {\it
  kglobal}. The results from these simulations are shown in
Fig.~\ref{tpar}. The PIC simulation is in solid black, {\it kglobal}
is in red, and the result from {\it kglobal} without a parallel
electric field is in dashed red. We added the latter so we could
determine how the addition of the parallel electric field influenced
the dynamics. First, the temperature profiles from {\it p3d} and {\it
  kglobal} with $E_\parallel$ match very well over most of the
domain. In contrast, the temperature in {\it kglobal} with
$E_\parallel =0$ spreads much more rapidly, demonstrating that
$E_\parallel$ does inhibit electron thermal transport and that the the
model for $E_\parallel$ in {\it kglobal} correctly describes transport
suppression.

\begin{figure}[h]
    \centering
    \includegraphics[width=16pc,height=8pc]{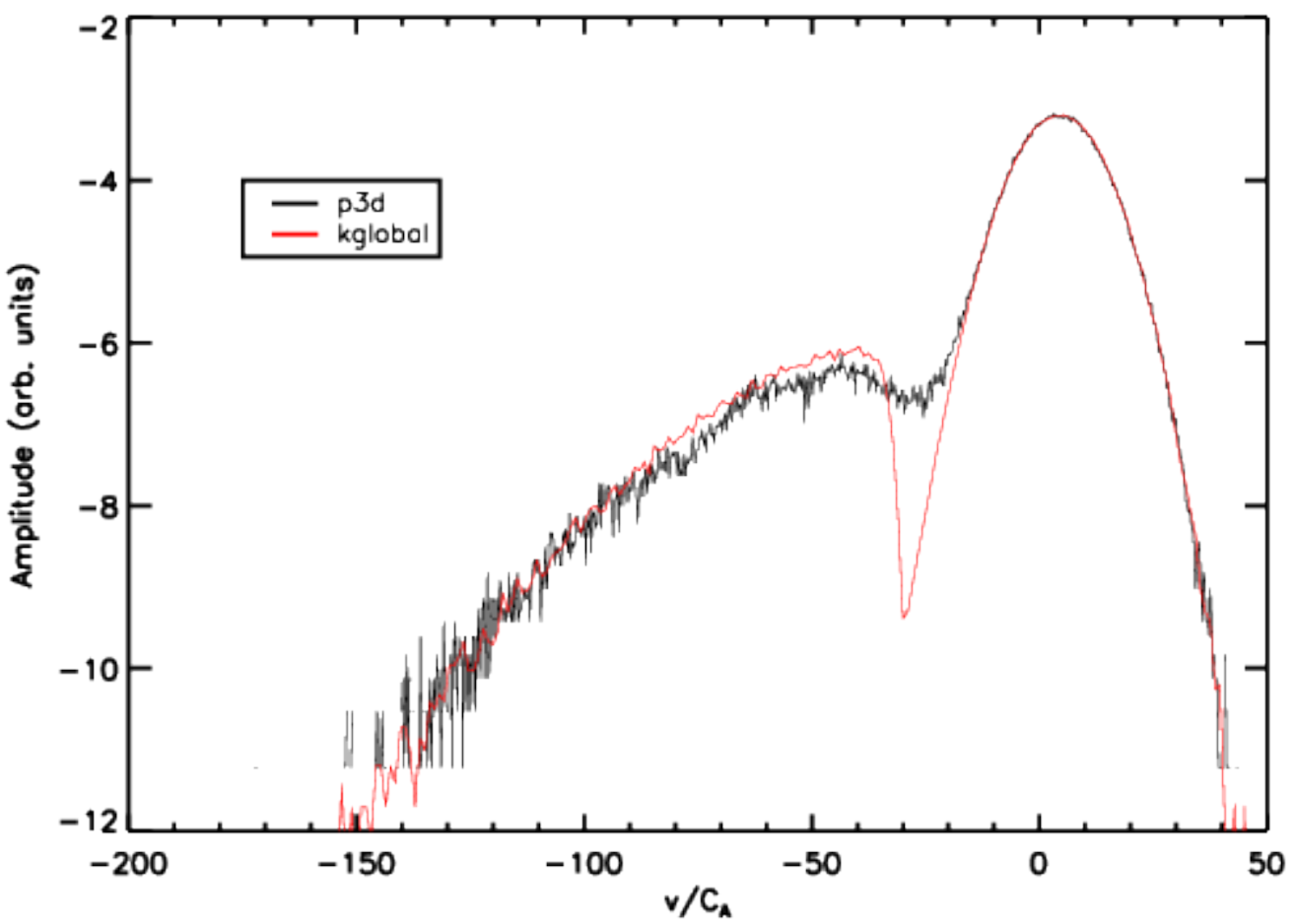}
    \caption{The log of the particle distribution functions from
      the PIC code {\it p3d} (black) and the {\it kglobal} code (red) taken
      at $x/L_0=0.25$. Notice the dip in the {\it kglobal}
      distribution function around $v/C_A=-30$.}
    \label{distr}
\end{figure}

While {\it kglobal} is able to capture the overall dynamics of the
temperature profile, it does
not produce the short scale spatial oscillations seen in the {\it p3d}
data. These oscillations are plasma waves driven unstable by a
bump-on-tail velocity distribution that smooths out the plateaus in
the temperature visible in the {\it kglobal} data around $0.3<x/L_0<0.4$
and $0.6<x/L_0<0.7$. Fig.~\ref{distr} displays the distribution
functions from {\it p3d} in black and from {\it kglobal} in red at
$x/L_0=0.25$ at the time shown in Fig.~\ref{tpar} panel (b). Note that for {\it
  kglobal} a Maxwellian with density and temperature equal to that of
the cold electron fluid was added to the hot electron distribution
function so that we could directly compare cuts to {\it p3d}. There is
a sharp dip visible in the velocity distribution from {\it kglobal}
around $v/C_A=-30$ that is not seen in the data from {\it p3d}. In the
{\it p3d} simulation, the faster particles have lost energy to plasma
oscillations and filled in this dip, forming a plateau in phase space. This result is not seen in the {\it kglobal} data
because this model does not support plasma waves, which require a
violation of charge neutrality to exist. Electron sound waves can be
driven unstable by structures in velocity space, but the phase speed of
these waves is fixed by the local plasma parameters (see
Eq.~(\ref{dispersion})) and so will typically not be resonant with
electrons in the bump region shown in Fig.~\ref{distr}.

\section{Conclusion}
The {\it kglobal} code \citep{Drake2019} has been upgraded to include a
macroscale $E_\parallel$ that develops as a result of gradients in the
plasma pressure parallel to the ambient magnetic field. The upgraded
model now captures the dynamics of electron acoustic waves and
accurately describes the suppression of transport of hot electrons
parallel to the ambient magnetic field, a process that is important in the
early phases of electron acceleration in magnetic reconnection \citep{Egedal2015, Haggerty2015}. The inclusion of the large scale $E_\parallel$ is also 
important in describing the development of return currents that form
as hot electrons escape from regions of electron acceleration in
macroscale energy release events such as flares in the solar corona.
This new capability combined with the ability of the model to describe
the impact of pressure anisotropy on magnetic field dynamics (e.g.,
firehose instability), which is critical for describing the feedback of
energetic particles on reconnection dynamics, suggest that the {\it
  kglobal} code can be used to accurately simulate nonthermal electron
acceleration during magnetic reconnection.

Our next step is to begin to explore the energization of electrons
during magnetic reconnection with {\it kglobal} and to determine
whether the reconnection dynamics in a macroscale system can produce
the power law distributions that are ubiquitous in observations
\citep{Lin2003,Gary2018}. Because {\it kglobal} is a macroscale model, the
dynamics of particle acceleration can be explored in a much larger
domain than with a traditional PIC model. In addition, we will include
particle loss in a realistic manner to establish whether or not it
is the balance between reconnection drive and the escape of energetic
particles that leads to powerlaw distributions \citep{Drake2013, Guo2014}. Finally, in a macroscale simulation model the inclusion of a
synthetic diagnostic to describe synchrotron emission and
bremsstrahlung emission will be possible.

\begin{acknowledgments}
 We would like to thank B. Daughton for invaluable discussions that led to this work.
 This work has been supported by NSF Grant Nos. PHY1805829 and
 PHY1500460 and the FIELDS team of the Parker Solar Probe (NASA
 contract NNN06AA01C). Joel Dahlin was supported by an appointment to the NASA Postdoctoral
Program at the NASA Goddard Space Flight Center, administered by Universities Space 
Research Association under contract with NASA. The simulations were carried out at the
 National Energy Research Scientific Computing Center (NERSC). The
 data used to perform the analysis and construct the figures for this
 paper are preserved at the NERSC High Performance Storage System and
 are available upon request.
\end{acknowledgments}
\appendix*
\section{Energy Conservation}

We start with the momentum equations for the three species, the ions
\begin{equation}
  \rho\frac{d\boldsymbol{v}_i}{dt}=ne\boldsymbol{E}+\frac{ne}{c}\boldsymbol{v}_i\times\boldsymbol{B}-\boldsymbol{\nabla}P_i,
  \label{eqn:mom_i}
\end{equation}
the cold electrons
\begin{multline}
\frac{\partial (m_en_cv_{\parallel c}\boldsymbol{b})}{\partial t}=-n_{c}e\boldsymbol{E}-\frac{n_{c}e}{c}\boldsymbol{v}_{c}\times\boldsymbol{B}-\boldsymbol{\nabla}P_{c} \\ -\boldsymbol{b}\boldsymbol{B}\cdot \boldsymbol{\nabla}\frac{m_en_cv_{\parallel c}^2}{B} - m_en_cv_{||c}^2 \boldsymbol{\kappa}
    \label{eqn:mom_c}
\end{multline}
and the hot electrons
\begin{equation}
\frac{\partial (m_en_{h}\bar{v}_{\parallel h}\boldsymbol{b})}{\partial t}=-n_{h}e\boldsymbol{E}-\frac{n_{h}e}{c}\bar{\boldsymbol{v}}_{h}\times\boldsymbol{B}-\boldsymbol{\nabla}\cdot\boldsymbol{T}_{h},
    \label{eqn:mom_h}
\end{equation}
where $\rho$ and $\boldsymbol{v}$ are the ion mass density and velocity, from charge neutrality $n=n_c+n_h$ and the electron inertia has only been retained in the direction
along the ambient magnetic field. In writing the electron momentum equations we have for simplicity assumed that the mean drifts of both species are not relativistic.  The individual electron fluxes can be of order $nC_{Ae}$ while the ion flux is of order $nC_A$. However, we show below that the total current is much smaller than the contribution from each species of particle and this yields a constraint on the total driver of the current. To see this we divide the momentum equations along the field lines by their respective masses and subtract Eqs.~(\ref{eqn:mom_c}) and (\ref{eqn:mom_h}) from Eq.~(\ref{eqn:mom_i}), which yields
\begin{multline}
\frac{1}{e} \frac{\partial J_{||}}{\partial t}=\frac{n_i e E_{||}}{m_e} - \boldsymbol{b} \cdot \left(\frac{1}{m_e} \boldsymbol{\nabla} P_c + \frac{1}{m_e} \boldsymbol{\nabla} \cdot \boldsymbol{T_h}\right) \\ - \boldsymbol{B} \cdot \boldsymbol{\nabla} \left( \frac{n_cv_{||c}^2}{B} \right)
\label{summom}
\end{multline}
All of the terms on the right hand side of this equation act as drivers of $J_{\parallel}$. However, the parallel current driven is constrained by the structure of the magnetic field which is produced by this current. This constraint follows from Amp\`ere's law $J_\parallel\sim cB/4\pi L$, where $L$ is the macroscopic characteristic perpendicular scale of the magnetic field. Comparing the time derivative of this current, given by $c_A/L$, with the characteristic scaling of the terms on the right, {\it e.g.}, the gradient of the hot thermal electrons, which scales as $n_hT_h/m_eL$, we find that the ratio of the left to the right side of the equation scales like $\sqrt{m_e/m_i}(d_e/L)\ll 1$. Thus, the time derivative of the current can be discarded. This tells us that
\begin{equation}
  v_{||c}=\frac{1}{n_c}(n_iv_{||i}-n_hv_{||h}).
  \label{v||c}
\end{equation}
Note that this constraint equation for $v_{||c}$ includes the ion motion. That the ions must also be included in the constraint follows because the mean drift speed associated with the current (from the previous scaling for $J_\parallel$) scales like  $nC_A(d_i/L)\ll nC_A$, the characteristic current carried by the ions. This constraint on the parallel flows is consistent with the conclusions of \citet{Kulsrud1983} and yields the equation for $E_\parallel$ in Eq.~(\ref{epar}). If the mean flows of the electrons becomes relativistic, corrections to Eq.~(\ref{epar}) of order $v_{\parallel h}/c$ must be included.

A further consequence of this result is that the sum of the fluxes of the two electron species is limited to a scale of the order of the ion flux. The consequence is that when the
three momentum equations are summed, the electron inertia arising from
the time derivative can be discarded, which yields the ion momentum equation,
\begin{multline}
  \rho\frac{d\boldsymbol{v}}{dt}=\frac{1}{c}\boldsymbol{J}\times\boldsymbol{B}-\boldsymbol{\nabla}(P_i+P_c)-\boldsymbol{\nabla }\cdot\boldsymbol{T}_{eh}-\boldsymbol{b} \boldsymbol{B}\cdot \boldsymbol{\nabla}\frac{m_en_cv_{\parallel c}^2}{B}\\-m_en_cv_{||c}^2\boldsymbol{\kappa},
  \label{eqn:mhd}
\end{multline}
which is equivalent to the form shown in Eq.~(\ref{fluidmom}). To explore energy conservation of Eqs.~(\ref{fluidmom}) and (\ref{eqn:ehot}) along with the usual fluid equations, we take the dot product of Eq.~(\ref{fluidmom}) with $\boldsymbol{v}$ and use the ion continuity equation to obtain
\begin{multline}
  \frac{\partial }{\partial t}\frac{\rho v^2}{2}+\boldsymbol{\nabla}\cdot\frac{\rho \boldsymbol{v} v^2}{2}+\boldsymbol{v}\cdot \boldsymbol{\nabla}P_i=(\boldsymbol{J}_\perp-\boldsymbol{J}_{\perp c}-\boldsymbol{J}_{\perp h})\cdot \boldsymbol{E}_\perp \\ -(J_{\parallel c}+J_{\parallel h})E_\parallel=\boldsymbol{J}_\perp\cdot\boldsymbol{E}_\perp-(\boldsymbol{J}_c+\boldsymbol{J}_h)\cdot \boldsymbol{E},
    \label{eqn:energy}
\end{multline}
where we have used the perpendicular components of Ohm's law $\boldsymbol{
  E}_\perp=-\boldsymbol{v}\times \boldsymbol{B}/c$, the perpendicular components of the two electron momentum equations and Eq.~(\ref{v||c}) for $v_{||c}$.  From Faraday's law we find
\begin{equation}
  \frac{\partial }{\partial t}\frac{B^2}{8\pi}+\frac{c}{4\pi} \boldsymbol{\nabla}\cdot (\boldsymbol{E}\times \boldsymbol{B})+\boldsymbol{J}_\perp\cdot \boldsymbol{E}_\perp=0,
\end{equation}
which, when combined with Eq.~(\ref{eqn:energy}), yields the conservation law
\begin{equation}
    W_{MHD}+W_c+W_h=constant,
\label{eqn:energycons}
\end{equation}
where we have discarded terms corresponding to the divergence of the various energy fluxes. The MHD energy, $W_{MHD}$, includes the ion bulk kinetic and thermal energies and the magnetic energy, the cold electron energy includes both the kinetic energy associated with parallel streaming and the thermal energy,
\begin{equation}
  W_c=\frac{m_en_cv_{\parallel c}^2}{2}+\frac{1}{\Gamma -1}P_c
  \label{eqn:wc}
  \end{equation}
with $\Gamma$ the ratio of specific heats. The hot electron energy is the sum of the parallel kinetic energies of all hot electrons as well as the energy associated with their perpendicular gyro motion. It does not include the kinetic energy associated with the perpendicular bulk flow, which is negligible.  

{}
\end{document}